\documentclass[twocolumn,groupedaddress,superscriptaddress,showpacs,letterpaper]{revtex4}

\usepackage{amsmath,amssymb}	
\usepackage{subfigure,graphicx} 
\usepackage{bm} 

\usepackage{hyperref}
\usepackage{times}
\hypersetup{%
  breaklinks = {true}, 
  citecolor = {blue},
  colorlinks = {true}, 
  linkbordercolor = {1 0 0},
  linkcolor = {red},
  pdfauthor = {\textcopyright\ Vitor M. Pereira},
  pdfcreator = {\LaTeX\ and \flqq hyperref\frqq},
  pdffitwindow = {true}, 
  pdfmenubar = {true},
  pdfpagelayout = {SinglePage},
  pdfstartview = {Fit}
  pdftitle = {Disorder Induced Localized States in Graphene},
  pdftoolbar = {true},
  plainpages = {false}, 
}

\begin{document}



\title{Disorder Induced Localized States in Graphene}


\date{\today}

\author{Vitor M. Pereira}
\affiliation{Department of Physics, Boston University, 590 
Commonwealth Avenue, Boston, MA 02215, USA}
\affiliation{CFP and Departamento de F{\'\i}sica, Faculdade de Ci\^encias
Universidade de Porto, 4169-007 Porto, Portugal}

\author{F. Guinea}
\affiliation{Department of Physics, Boston University, 590 
Commonwealth Avenue, Boston, MA 02215, USA}
\affiliation{Instituto de  Ciencia de Materiales de Madrid, CSIC,
 Cantoblanco E28049 Madrid, Spain}

\author{J.~M.~B. Lopes dos Santos}
\affiliation{CFP and Departamento de F{\'\i}sica, Faculdade de Ci\^encias
Universidade de Porto, 4169-007 Porto, Portugal}

\author{N.~M.~R. Peres}
\affiliation{Department of Physics, Boston University, 590 
Commonwealth Avenue, Boston, MA 02215, USA}
\affiliation{Center of Physics and Departamento de F{\'\i}sica,
Universidade do Minho, P-4710-057, Braga, Portugal}

\author{A.~H. Castro Neto}
\affiliation{Department of Physics, Boston University, 590 
Commonwealth Avenue, Boston, MA 02215, USA}


\begin{abstract}
We consider the electronic structure near vacancies in the 
half-filled honeycomb lattice.  
It is shown that vacancies induce the formation of localized states. 
When particle-hole symmetry is broken, localized states become 
resonances close to the Fermi level. We also study the problem 
of a finite density of vacancies, obtaining the electronic
density of states, and discussing the issue of electronic 
localization in these systems. 
Our results have also relevance for the problem of disorder 
in d-wave superconductors.
\end{abstract}

\maketitle


{\it Introduction}.
The problem of disorder in systems with Dirac fermions has been
studied extensively in the last few years in the context of 
dirty d-wave superconductors \cite{HA02}. Dirac fermions are
also the elementary excitations of the honeycomb lattice at half-filling,
equally known as graphene, which is realized in two-dimensional (2D) Carbon 
based materials with sp$^2$ bonding. It is well-known that disorder
is ubiquitous in graphene and graphite (which is produced by stacking
graphene sheets) and its effect on the electronic structure has 
been studied extensively
\cite{OS91,GGV92,CEL96,WS00,W01,GGV01,H01,MA01,Hetal04,DSL04,Letal04,Vetal05}.
It has been shown recently \cite{PGC05} that the interplay of disorder and
electron-electron interactions is fundamental for the understanding 
of recent experiments in graphene devices \cite{Netal04}.
Furthermore, experiments reveal
that ferromagnetism is generated in heavily disordered graphite samples
\cite{Eetal02,Eetal03b,Retal04,MP05,Metal05}, but the understanding 
of the interplay of strong disorder and electron-electron interactions in
these systems is still in its infancy. Different mechanisms for
ferromagnetism in graphite have been proposed and they are either 
based on the nucleation of ferromagnetism around extended defects 
such as edges \cite{OS91,H01,Letal04,Vetal05} or due to exchange
interactions originating from unscreened Coulomb interactions \cite{nuno2}.
Therefore, the understanding of the nature of the electronic states
in Dirac fermion systems with strong disorder is of the utmost interest. 

In the following, we analyze in detail states near the Fermi energy 
induced by vacancies in a tight-binding model for the electronic states 
of graphene planes. We show that single vacancies in a graphene plane
generate localized states which are sensitive to the presence of
particle-hole symmetry breaking. Moreover, a finite density
of such defects leads to strong changes in the local and averaged 
electronic Density Of States (DOS) with the creation of localized states
at the Dirac point. 


{\it The model.}
We consider a single band model described by the Hamiltonian:
\begin{equation}
{\cal H} = - t \sum_{\langle i,j \rangle} c^\dag_i c_j  - t' 
\sum_{\langle \langle i,j \rangle \rangle} c^\dag_i c_j + {\rm h.c.} \, ,
\label{hamil}
\end{equation} 
where $c_j$ ($c_j^\dag$) annihilates (creates) an electron at 
site ${\bf R}_j$ of the honeycomb lattice (the spin quantum numbers
are suppressed since we do not consider spin dependent phenomena).
In (\ref{hamil}) $t$ is the nearest neighbor hopping energy  
($t \approx 2.7$eV in graphene) and $t'$ is the next-nearest neighbor
hopping energy. In the honeycomb lattice, while $t$ describes the hopping
of electrons between the two sub-lattices, 
$t'$ describes the hopping in the same sub-lattice and therefore 
breaks particle-hole symmetry. The  relevance of $t'$ for graphene 
is suggested by different experiments ($t' \approx 0.2 t$) \cite{PGC05}. 

\begin{figure}[t]
\begin{center}
\includegraphics*[width=1.0\columnwidth]{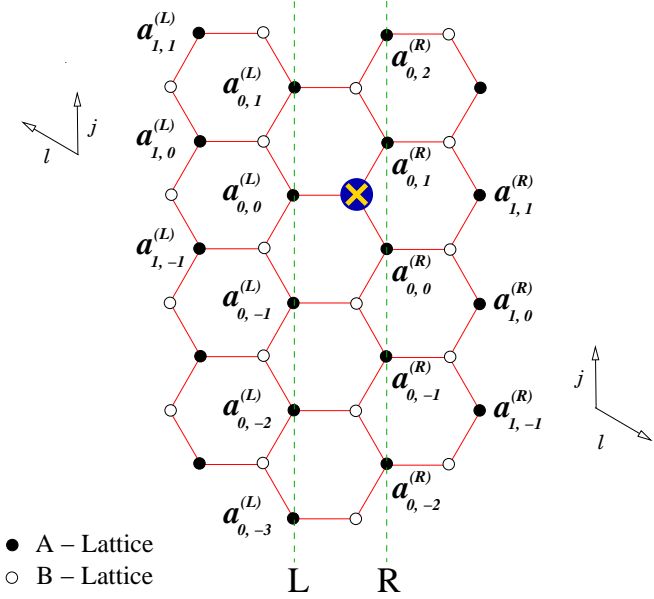}
\end{center}
\caption{(color online) Sketch of the geometry considered in the text for the study of a
  single B-site vacancy ($\otimes$). The system has periodic boundary
  conditions along the vertical axis, and is infinite along the horizontal
  axis. Only some A-site amplitudes are shown.}
\label{sketch_vacancy}
\end{figure}

{\it Localized states around vacancies.}
Firstly we consider the particle-hole symmetric case ($t'=0$), and 
use the geometry shown in Fig.~\ref{sketch_vacancy}. We analyze a
cluster with periodic boundary conditions along the vertical direction, like
a zigzag nanotube. In the absence of vacancies, the Hamiltonian can be
simplified using the translational symmetry along the vertical direction. 
The states can be classified by the momentum (in units of $1/a$,
with $a$ the lattice constant) along the vertical
axis, $k_m = ( 2 \pi m )/N \, , m = 1,..., N$ where $N$ is the number of
unit cells along the vertical axis. The system is metallic if $N = 3M$,
where $M$ is an integer. Using this set of momenta, the wavefunction amplitudes
in each sub-lattice ($A$,$B$) can be written as
$a_{l,j}=\sum_{k_m}a_{l,k_m} e^{ik_m j}$ and 
$b_{l,j}=\sum_{k_m} b_{l,k_m} e^{i k_m j}$, where $(l,j)$ are the 
unit cell coordinates in the geometry of Fig.~\ref{sketch_vacancy}. 
Upon this transformation, the original problem is mapped, for each 
value of $k_m$, into a one-dimensional (1D) tight-binding model along 
the horizontal direction, with two sites per unit cell. 
Such 1D problem is characterized by site amplitudes $a_{l} \equiv a_{l,k_m}$ 
and $b_{l} \equiv b_{l,k_m}$, and by two effective hoppings, $t$ and 
$t ( 1+ e^{i k_m} )$; a  gauge transformation allows us to make 
them both real, reading $t$ and $2 t \cos ( k_m / 2 )$.

%
%

For the solution of the impurity problem, it is convenient to
define two planes contiguous to the vacancy, as shown in
Fig.~\ref{sketch_vacancy}. The two planes belong to the same 
sub-lattice, opposite to the one where the vacancy resides. 
A possible localized state at zero energy must: 
({\it i}) decay as one moves along the horizontal axis away from 
these planes, ({\it ii}) satisfy the bulk tight binding 
equations arising from the eigenvalue condition 
$H | \Psi \rangle = E | \Psi \rangle = 0$
at these axes and beyond, and ({\it iii}) have 
amplitudes $a^{(L)}_{0,j}$ and $a^{(R)}_{0,j}$ which satisfy the equations,
\begin{equation}
a^{(L)}_{0,j} = - a^{(R)}_{0,j} - a^{(R)}_{0,j+1}, 
\label{matching}
\end{equation} 
for all $j$, except at the vacancy ($j=0$). 
Having $b^{(L,R)}_{l,j}=0$ everywhere automatically satisfies 
condition ({\it ii}).
Plane $L$ at the left of the vacancy can be considered a zig-zag edge. 
To its left, one can define states
which decay exponentially for momenta such that
$ 2\pi/3 \le k_m \le 4 \pi /3$ %
\cite{WS00,W01}. The associated wavefunctions decay as
$\vert 2 \cos ( k_m / 2 ) \vert^l$. As defined
in Fig.~\ref{sketch_vacancy}, $l$ is always positive, since
it grows away from both the $L$ and the $R$ planes, and can 
be interpreted as the distance to plane $L$ divided by $3a/2$. 
Analogously, one can define localized wavefunctions to 
the right of plane $R$, for momenta such that
$ 0 \le k_m \le 2\pi / 3$ or 
$4 \pi / 3 \le k_m \le 2 \pi$.
These wavefunctions decay as 
$| 2 \cos ( k_m / 2 ) |^{- l}$.
The amplitudes $a^{(L)}_{0,j}$ and $a^{(R)}_{0,j}$ at planes 
$L$ and $R$ can be written, in terms of momentum eigenstates, as:
\begin{eqnarray}
a^{(L)}_{0,j} = \sum_{k_m} a^{(L)}_{k_m} e^{i k_m j} \, ,\hspace{1cm}
a^{(R)}_{0,j} = \sum_{k_{m'}} a^{(R)}_{k_{m'}} e^{i k_{m'} j} \,,
\label{amplitudes}
\end{eqnarray}
and if there was no impurity present in the system, we would have
$k_{m}$ and $k_{m'}$ in the range $[0,2\pi[$.
The boundary condition introduced by Eq. (\ref{matching}) can be written as:
\begin{equation}
\sum_{k_m} a^{(L)}_{k_m} e^{i k_m j} = - \sum_{k_{m'}} \left( 1 + e^{i k_{m'}}
\right) a^{(R)}_{k_{m'}} e^{i k_{m'} j} \, ,
\end{equation}
for $j \neq 0$. These equations admit the solution:
\begin{eqnarray}
a^{(L)}_{k_m} = 1 \, ,\hspace{1cm}
a^{(R)}_{k_{m'}} \left( 1 + e^{i k_{m'}} \right) = 1 \, ,
\end{eqnarray}
where the wavevectors $k_m$ and $k_{m'}$ satisfy 
$2 \pi/3 \le k_m \le 4 \pi /3$ and $0 \le k_{m'} \le 2\pi / 3$ or 
$4 \pi / 3 \le k_{m'} \le 2 \pi$. We note that these
momentum values are the same defining the decaying states
to the left and to the right of planes $L$ and $R$, respectively, 
if we had considered the two cases as separate problems. Hence, 
we can use the amplitudes we found in the latter case (given above)
when constructing the wave function for an impurity.
To the left of plane $L$, the 
amplitude $a^{(L)}_{l,j}$ is now given as 
$a^{(L)}_{l,j}=\sum_{k_m} \left[-2\cos(k/2)\right]^l \exp[ik_m(j+l/2)]$,
where the values of $k_m$ are those imposed by the boundary condition,
and a similar expression for $a^{(R)}_{l,j}$.
For sufficiently wide nanotubes (corresponding to the solution for 
the infinite lattice) we can approximate the sums in $k_m$ by
integrals. Shifting from lattice position coordinates to distances 
relative to the $L$ plane, the amplitude $a^{(L)}_{l,j}$ gives the 
wavefunction $\Psi ( x=l3a/2 , y=a\sqrt3(j+l/2) )$ 
at a point  of coordinates $(x,y)$.
In units of the lattice constant, $\Psi ( x , y )$  is approximately
\begin{eqnarray}
\Psi^{(L)}(x,y) & \sim & \int_{2\pi/3}^{4\pi/3} dk      %
                   \left( -2 \cos(k/2) \right)^{2x/3}   %
                    e^{i k y/\sqrt{3}} \nonumber \\ 
                & \approx & \frac{e^{(4\pi iy)/(3\sqrt{3})}}{x+iy} + %
                         \frac{e^{ 2\pi i (x+y/\sqrt3) / 3}} {x-iy} \, ,
\label{psi_x_y}
\end{eqnarray}
when the lattice site $( x, y )$ is in the  opposite sub-lattice of the
vacancy, and $\Psi ( x, y ) = 0$ when $( x, y )$ is in  
the same sub-lattice as the vacancy. It is now clear
that extra vacancies added to the sub-lattice where the impurity resides, 
cannot change this wave function.

\begin{figure}[t]
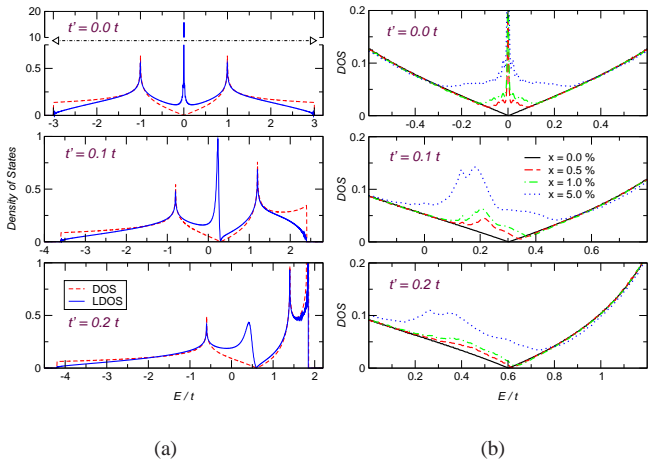

\begin{center}
\subfigure[]{\includegraphics*[width=0.49\columnwidth]{Fig_2a.eps}
\label {single_vacancy}}
\subfigure[]{\includegraphics*[width=0.48\columnwidth]{Fig_2b.eps}
\label{many_vacancies}}
\end{center}
\caption{(color online) \subref{single_vacancy} 
        Comparison between the local DOS in the vicinity of a
        vacancy (blue/continuous) with the bulk DOS (red/dashed) 
        in clean systems. 
        \subref{many_vacancies} 
        Total DOS in the vicinity of the Dirac 
        points for clusters with $4\times10^6$ sites, at selected 
        vacancy concentrations. Numerical results in (a,b) obtained for $t'=0$ 
        (top panels), $t'=0.1\,t$ (center) and $t'=0.2\,t$ (bottom). 
				(Notice the scale truncation in the upper part of the first 
         panel in \subref{single_vacancy}.}
\end{figure}

We would like to understand this results, from the point of view
of a low-energy effective theory.
The eigenstates of the discrete Hamiltonian (\ref{hamil}) can be
approximated, at long wavelengths, by the Dirac equation.
The wavefunctions can be written as a spinor:
\begin{equation}
\Psi ( x , y ) \equiv \left( \begin{array}{c} \psi^{a,b}_1 ( x , y ) \\ 
\psi^{a,b}_2 ( x  , y ) \end{array} \right)
\label{wavefunction}
\end{equation}
where the functions $\psi_1 ( x , y )$ and $\psi_2 ( x , y )$ correspond to
the amplitudes of the wavefunctions in each of the two sub-lattices. There are
two set of spinors, $a,b$, which correspond to the two inequivalent states at
the corners of the Brillouin zone. At zero energy, the functions 
at one of the Dirac points satisfy:
$
\partial_z \psi^a_1 ( z , \bar{z} ) = 0 \, ,
\partial_{\bar{z}} \psi^a_2 ( z , \bar{z} ) = 0 \,%
$,
where $z = x + i y$ and $\bar{z} = x - i y$;
and those at the other Dirac point can be obtained by replacing $z$ by
$\bar{z}$ everywhere. The result of Eq. (\ref{psi_x_y}) implies that the
boundary conditions at the vacancy are such that the combination
\begin{equation}
\Psi ( x , y ) = \left( \begin{array}{c} \psi^a_1 ( z ) + \psi^b_1 ( 
\bar{z} ) \\ 0  \end{array} \right) \propto \left( \begin{array}{c} 
\frac{1}{z} + \frac{1}{\bar{z}} \\ 0 \end{array} \right) \, 
\label{dirac_sol}
\end{equation} 
is selected (notice that, in the long wavelength limit, the phase factors 
in Eq.~(\ref{psi_x_y}) are factored out upon defining the slowly 
varying Dirac fields). 
This solution, although decaying away from the impurity, is not normalizable. 
It has the same spatial dependence as the quasi-localized
solutions which are induced by radial potentials on 2D 
Dirac fermions \cite{DHM98}. 
The matching of localized states described above cannot
be generalized to the case $t' \ne 0$, as the band of edge states is not
degenerate in energy \cite{Mudry:2002}. The localized state at the Fermi
level becomes a resonance inside the continuum of extended states.
Numerical results for the local (DOS) at a site near a single
vacancy are shown in Fig.~\ref{single_vacancy}. In the absence of
electron-hole symmetry, the localized state becomes a resonance with
increasing width, shifted from the Fermi energy.

%
%

The scheme used here can be generalized to study other lattice
representations of the Dirac equation with electron-hole symmetry. The
electronic structure near impurities in d-wave superconductors has been
studied using the Hamiltonian:
 \cite{Aetal00,AHM00,AHZ03}: 
\begin{equation}
{\cal H}_{\rm BCS} = t \sum_{\langle i,j \rangle} c^\dag_i c_j + \Delta \left(
  c^\dag_{i,j} c^\dag_{i\pm {\bf a},j} - c^\dag_{i,j} c^\dag_{i.j \pm {\bf
  b}} \right) + {\rm h. c.} \, ,
\label{hamil_BCS}
\end{equation}
where the sites are defined in a square lattice, ${\bf a}$ and ${\bf b}$ are
the lattice vectors along the horizontal and vertical directions. 
The Hamiltonian (\ref{hamil_BCS}) is formally identical to a tight-binding 
model on a square lattice and two orbitals per site. The hopping
terms between different orbitals have different signs along the horizontal
and vertical axis. If periodic boundary conditions are applied along the
$(1,1)$ and $(1,-1)$ directions, the problem can be written as the matching
of a set of 1D wavefunctions, in a way similar to the scheme depicted in
Fig.~\ref{sketch_vacancy}. A vacancy leads to a quasi-localized state,
described by the same long wavelength wavefunction (see also \cite{Ketal97}). 
The quasi-localized state described here, being localized in a single
sub-lattice, remains a solution when there are two vacancies in different
sub-lattices, in agreement with numerical calculations \cite{ZAH03}.


{\it Finite densities of vacancies}.
We have extended the previous results to systems with a finite
density of vacancies, using the stochastic recursion method to obtain 
the DOS in clusters with up to $\sim10^6$ sites. 
Results for different impurity concentrations and different values of $t'$
are shown in Fig.~\ref{many_vacancies}. In the presence of electron-hole
symmetry ($t'=0$), the inclusion of vacancies brings an increase of spectral 
weight to the surroundings of the Dirac point, leading to a DOS whose 
behavior for  $E\approx0$ mostly resembles the results obtained within a 
Coherent Potential Approximation (CPA) \cite{PGC05}. The most important 
feature,  however, is the emergence of a sharp peak at the Fermi level, 
superimposed upon the flat portion of the DOS (apart from the peak, the 
DOS flattens out in this neighborhood as $x$ is increased past the $5\,\%$ 
shown here). 
The breaking of the particle-hole symmetry by a finite $t'$ results in the 
broadening of the peak at the Fermi energy, and the displacement of its 
position by an amount of the order of $t'$. All these effects take place 
close to the the Fermi energy. At higher energies, the only deviations from 
the DOS of a clean system are the softening of the van~Hove singularities 
and the development of Lifshitz tails (not-shown) at the  band edge, 
both induced by the 
increasing disorder caused by the random dilution. The onset of this high 
energy regime, where the profile of the DOS is essentially unperturbed by 
the presence of vacancies, is determined by 
$\epsilon \approx v_{\rm F} / l$, $l \sim n_{\rm imp}^{-1/2}$ being the 
average distance between impurities.

\begin{figure}[t]
\begin{center}
\includegraphics*[width=0.8\columnwidth]{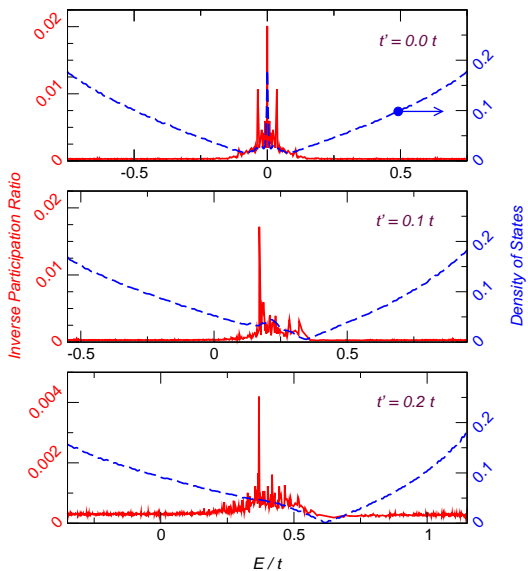}
\end{center}
\caption{(color online)
        Inverse participation ratio for a concentration of vacancies of 0.5~\% in 
        systems with $10^4$ sites. The dashed curve is the total DOS for the same 
				dilution. Top: $t'=0$; center: $t'=0.1\,t$;  bottom: $t'=0.2\,t$. 
        Only the low energy region is shown.}
\label{part_ratio}
\end{figure}

To address the degree of localization for the states near the Fermi
level, the Inverse Participation Ratio (IPR) was calculated via exact 
diagonalization on smaller systems. 
For an eigenstate $m$, the IPR is the quantity defined as:
${\cal P}_m  = \sum_i^N | \Psi_m (  i ) |^4 $,
the index $i$ labeling the lattice sites. The wavefunction of an 
extended state has an amplitude equally significant throughout the 
entire system ($\Psi_m(i) \sim N^{-1/2}$), whence we naturally expect 
${\cal P}_m \sim N^{-1}$. For a localized state, in opposition, 
its very definition entails the fact that only a finite number of 
lattice sites will contribute to the normalization, resulting 
in much higher values of ${\cal P}_m $. Results for different 
values of $t'$ are shown in Fig.~\ref{part_ratio} for random dilution 
at $0.5\,\%$. One observes, first, that ${\cal P}_m \sim 3/N$ for 
all energies but the Fermi level neighborhood, as expected for states
extended up to the length scale of the system sizes used in the numerics. 
Secondly, the IPR becomes significant exactly in the same energy 
range where the DOS exhibits the vacancy-induced anomalies discussed above. 
Clearly, the farther the system is driven from the particle-hole 
symmetric case, the weaker the localization effect, as illustrated 
by the results obtained with $t'=0.2\,t$. To this respect, it is 
worth mentioning that the magnitude of the strongest peaks in 
${\cal P}_m$ at $t'= 0$ and $t'=0.1\,t$ is equal to the magnitude 
of the IPR calculated for a single impurity problem 
\footnote{For a single impurity, $\mathcal{P}_{E=0}$ is seen to 
vary with $N$ like $\log^{-2}(N)$, as expected for such 
quasi-localized state whose normalization depends upon the system size 
(cfr. Eqs.~\ref{psi_x_y}, \ref{dirac_sol}).}.
Such results indicate the existence of quasi-localized states at 
the center of the resonance, induced by the presence of the vacancies. 
For higher doping strengths, the enhancement of ${\cal P}_m$ is weaker 
in the regions where the DOS becomes flat, in agreement with the reasonably 
good description obtained with the CPA.

%
%


{\it Conclusions.}
We have studied the local DOS near vacancies in graphene
planes. The global and local DOS, are  
valuable tools  in the connection between a microscopic theory and the 
interpretation of local spectroscopic experiments, of which recent 
scanning tunneling spectroscopy measurements are an example \cite{Metal05}.
In agreement with general arguments valid for the 2D
Dirac equation, we find quasi-localized states at the Fermi level, if the
clean electronic structure shows electron-hole symmetry. In the absence of
electron-hole symmetry, these states are shifted and broadened. 
In a system with a finite concentration of impurities, we find a sharp peak
built up from localized states, superimposed on a \emph{structureless} finite
DOS, also induced by the vacancies. 
We have not considered here additional potentials or local atomic
displacements which can take place near vacancies. Our findings indicate that
the main features in the electronic structure are extended over many
lattice sites around the impurity. Thus, we expect that they will be weakly
affected by local modifications of the potential near the position of the 
vacancy. Finally, we have not discussed the implications of our results for the
magnetic properties of graphene planes. An enhancement of the density of
states over distances which are large compared to the lattice spacing, implies
that vacancies may lead to the formation of extended magnetic
moments, enhancing the tendency of the 
system towards ferro or antiferromagnetism.  


\begin{acknowledgements}
We thank C. Mudry for valuable discussions. 
J.M.B.L.S., N.M.R.P. and F.G. are thankful to the Quantum Condensed Matter
Visitor's Program at Boston University.
V.M.P. acknowledges the support of FCT, through grant SFRH/BD/4655/2001, 
and Boston University for the hospitality.
A.H.C.N. was partially supported through NSF grant DMR-0343790.
N.M.R.P. thanks FCT for partially supporting his sabbatical leave.
J.M.B.L.S. and V.M.P. were additionally financed by FCT and EU through
POCTI (QCAIII).
\end{acknowledgements}


\bibliography{vacancy}

\newcommand{\npb}{Nucl. Phys.}\newcommand{\adv}{Adv.
  Phys.}\newcommand{\epl}{Europhys. Lett.}
\begin{thebibliography}{28}
\expandafter\ifx\csname natexlab\endcsname\relax\def\natexlab#1{#1}\fi
\expandafter\ifx\csname bibnamefont\endcsname\relax
  \def\bibnamefont#1{#1}\fi
\expandafter\ifx\csname bibfnamefont\endcsname\relax
  \def\bibfnamefont#1{#1}\fi
\expandafter\ifx\csname citenamefont\endcsname\relax
  \def\citenamefont#1{#1}\fi
\expandafter\ifx\csname url\endcsname\relax
  \def\url#1{\texttt{#1}}\fi
\expandafter\ifx\csname urlprefix\endcsname\relax\def\urlprefix{URL }\fi
\providecommand{\bibinfo}[2]{#2}
\providecommand{\eprint}[2][]{\url{#2}}

\bibitem[{\citenamefont{Hirschfeld and Atkinson}(2002)}]{HA02}
\bibinfo{author}{\bibfnamefont{P.~J.} \bibnamefont{Hirschfeld}}
  \bibnamefont{and} \bibinfo{author}{\bibfnamefont{W.~A.}
  \bibnamefont{Atkinson}}, \bibinfo{journal}{J. Low Temp. Phys.}
  \textbf{\bibinfo{volume}{126}}, \bibinfo{pages}{881} (\bibinfo{year}{2002}).

\bibitem[{\citenamefont{Ovchinnikov and Shamovsky}(1991)}]{OS91}
\bibinfo{author}{\bibfnamefont{A.~A.} \bibnamefont{Ovchinnikov}}
  \bibnamefont{and} \bibinfo{author}{\bibfnamefont{I.~L.}
  \bibnamefont{Shamovsky}}, \bibinfo{journal}{Journ. of Mol. Struc. (Theochem)}
  \textbf{\bibinfo{volume}{251}}, \bibinfo{pages}{133} (\bibinfo{year}{1991}).

\bibitem[{\citenamefont{{J. Gonz\'alez \textit{et al.}}}(1992)}]{GGV92}
\bibinfo{author}{\bibnamefont{{J. Gonz\'alez \textit{et al.}}}},
  \bibinfo{journal}{\prl} \textbf{\bibinfo{volume}{69}}, \bibinfo{pages}{172}
  (\bibinfo{year}{1992}).

\bibitem[{\citenamefont{{J. C. Charlier \textit{et al.}}}(1996)}]{CEL96}
\bibinfo{author}{\bibnamefont{{J. C. Charlier \textit{et al.}}}},
  \bibinfo{journal}{Phys. Rev. B} \textbf{\bibinfo{volume}{53}},
  \bibinfo{pages}{11108} (\bibinfo{year}{1996}).

\bibitem[{\citenamefont{Wakabayashi and Sigrist}(2000)}]{WS00}
\bibinfo{author}{\bibfnamefont{K.}~\bibnamefont{Wakabayashi}} \bibnamefont{and}
  \bibinfo{author}{\bibfnamefont{M.}~\bibnamefont{Sigrist}},
  \bibinfo{journal}{\prl} \textbf{\bibinfo{volume}{84}}, \bibinfo{pages}{3390}
  (\bibinfo{year}{2000}).

\bibitem[{\citenamefont{Wakabayashi}(2001)}]{W01}
\bibinfo{author}{\bibfnamefont{K.}~\bibnamefont{Wakabayashi}},
  \bibinfo{journal}{\prb} \textbf{\bibinfo{volume}{64}},
  \bibinfo{pages}{125428} (\bibinfo{year}{2001}).

\bibitem[{\citenamefont{{J. Gonz\'alez \textit{et al.}}}(2001)}]{GGV01}
\bibinfo{author}{\bibnamefont{{J. Gonz\'alez \textit{et al.}}}},
  \bibinfo{journal}{\prb} \textbf{\bibinfo{volume}{63}},
  \bibinfo{pages}{134421} (\bibinfo{year}{2001}).

\bibitem[{\citenamefont{Harigaya}(2001)}]{H01}
\bibinfo{author}{\bibfnamefont{K.}~\bibnamefont{Harigaya}},
  \bibinfo{journal}{Journ. of Phys. C: Condens. Matt.}
  \textbf{\bibinfo{volume}{13}}, \bibinfo{pages}{1295} (\bibinfo{year}{2001}).

\bibitem[{\citenamefont{Matsumura and Ando}(2001)}]{MA01}
\bibinfo{author}{\bibfnamefont{H.}~\bibnamefont{Matsumura}} \bibnamefont{and}
  \bibinfo{author}{\bibfnamefont{T.}~\bibnamefont{Ando}}, \bibinfo{journal}{J.
  Phys. Soc. Japan} \textbf{\bibinfo{volume}{70}}, \bibinfo{pages}{2657}
  (\bibinfo{year}{2001}).

\bibitem[{\citenamefont{{K. Harigaya \textit{et al.}}}(2004)}]{Hetal04}
\bibinfo{author}{\bibnamefont{{K. Harigaya \textit{et al.}}}},
  \bibinfo{journal}{J. Phys. Chem. Sol.} \textbf{\bibinfo{volume}{65}},
  \bibinfo{pages}{123} (\bibinfo{year}{2004}).

\bibitem[{\citenamefont{{E. J. Duplock \textit{et al.}}}(2004)}]{DSL04}
\bibinfo{author}{\bibnamefont{{E. J. Duplock \textit{et al.}}}},
  \bibinfo{journal}{Phys. Rev. Lett.} \textbf{\bibinfo{volume}{92}},
  \bibinfo{pages}{225502} (\bibinfo{year}{2004}).

\bibitem[{\citenamefont{{P. O. Lehtinen \textit{et al.}}}(2004)}]{Letal04}
\bibinfo{author}{\bibnamefont{{P. O. Lehtinen \textit{et al.}}}},
  \bibinfo{journal}{Phys. Rev. Lett.} \textbf{\bibinfo{volume}{93}},
  \bibinfo{pages}{187202} (\bibinfo{year}{2004}).

\bibitem[{\citenamefont{{M. A. H. Vozmediano \textit{et al.}}}(2005)}]{Vetal05}
\bibinfo{author}{\bibnamefont{{M. A. H. Vozmediano \textit{et al.}}}}
  (\bibinfo{year}{2005}), \eprint{cond-mat/0505557}.

\bibitem[{\citenamefont{{N. M. R. Peres \textit{et
  al.}}}(2005{\natexlab{a}})}]{PGC05}
\bibinfo{author}{\bibnamefont{{N. M. R. Peres \textit{et al.}}}}
  (\bibinfo{year}{2005}{\natexlab{a}}), \eprint{cond-mat/0506709}.

\bibitem[{\citenamefont{{K. S. Novoselov \textit{et al.}}}(2004)}]{Netal04}
\bibinfo{author}{\bibnamefont{{K. S. Novoselov \textit{et al.}}}},
  \bibinfo{journal}{Science} \textbf{\bibinfo{volume}{306}},
  \bibinfo{pages}{666} (\bibinfo{year}{2004}).

\bibitem[{\citenamefont{{P. Esquinazi \textit{et al.}}}(2002)}]{Eetal02}
\bibinfo{author}{\bibnamefont{{P. Esquinazi \textit{et al.}}}},
  \bibinfo{journal}{\prb} \textbf{\bibinfo{volume}{66}},
  \bibinfo{pages}{024429} (\bibinfo{year}{2002}).

\bibitem[{\citenamefont{{P. Esquinazi \textit{et al.}}}(2003)}]{Eetal03b}
\bibinfo{author}{\bibnamefont{{P. Esquinazi \textit{et al.}}}},
  \bibinfo{journal}{Phys. Rev. Lett.} \textbf{\bibinfo{volume}{91}},
  \bibinfo{pages}{227201} (\bibinfo{year}{2003}).

\bibitem[{\citenamefont{{A. V. Rode \textit{et al.}}}(2004)}]{Retal04}
\bibinfo{author}{\bibnamefont{{A. V. Rode \textit{et al.}}}},
  \bibinfo{journal}{Phys. Rev. B} \textbf{\bibinfo{volume}{70}},
  \bibinfo{pages}{054407} (\bibinfo{year}{2004}).

\bibitem[{\citenamefont{Makarova and Palacio}(2005)}]{MP05}
\bibinfo{editor}{\bibfnamefont{T.}~\bibnamefont{Makarova}} \bibnamefont{and}
  \bibinfo{editor}{\bibfnamefont{F.}~\bibnamefont{Palacio}}, eds.,
  \emph{\bibinfo{title}{Carbon-Based Magnetism: an overview of metal free
  carbon-based compounds and materials}} (\bibinfo{publisher}{Elsevier,
  Amsterdam}, \bibinfo{year}{2005}).

\bibitem[{\citenamefont{{T. Matsui \textit{et al.}}}(2005)}]{Metal05}
\bibinfo{author}{\bibnamefont{{T. Matsui \textit{et al.}}}},
  \bibinfo{journal}{Phys. Rev. Lett.} \textbf{\bibinfo{volume}{94}},
  \bibinfo{pages}{226403} (\bibinfo{year}{2005}).

\bibitem[{\citenamefont{{N. M. R. Peres \textit{et
  al.}}}(2005{\natexlab{b}})}]{nuno2}
\bibinfo{author}{\bibnamefont{{N. M. R. Peres \textit{et al.}}}}
  (\bibinfo{year}{2005}{\natexlab{b}}), \eprint{cond-mat/0507061}.

\bibitem[{\citenamefont{{S.-H. Dong \textit{et al.}}}(1998)}]{DHM98}
\bibinfo{author}{\bibnamefont{{S.-H. Dong \textit{et al.}}}},
  \bibinfo{journal}{Phys. Rev. A} \textbf{\bibinfo{volume}{58}},
  \bibinfo{pages}{2160} (\bibinfo{year}{1998}).

\bibitem[{\citenamefont{{P. W. Brouwer \textit{et al.}}}(2002)}]{Mudry:2002}
\bibinfo{author}{\bibnamefont{{P. W. Brouwer \textit{et al.}}}},
  \bibinfo{journal}{Phys. Rev. B} \textbf{\bibinfo{volume}{66}},
  \bibinfo{pages}{14204} (\bibinfo{year}{2002}).

\bibitem[{\citenamefont{{W. A. Atkinson \textit{et
  al.}}}(2000{\natexlab{a}})}]{Aetal00}
\bibinfo{author}{\bibnamefont{{W. A. Atkinson \textit{et al.}}}},
  \bibinfo{journal}{Phys. Rev. Lett.} \textbf{\bibinfo{volume}{85}},
  \bibinfo{pages}{3926} (\bibinfo{year}{2000}{\natexlab{a}}).

\bibitem[{\citenamefont{{W. A. Atkinson \textit{et
  al.}}}(2000{\natexlab{b}})}]{AHM00}
\bibinfo{author}{\bibnamefont{{W. A. Atkinson \textit{et al.}}}},
  \bibinfo{journal}{Phys. Rev. Lett.} \textbf{\bibinfo{volume}{85}},
  \bibinfo{pages}{3922} (\bibinfo{year}{2000}{\natexlab{b}}).

\bibitem[{\citenamefont{{W. A. Atkinson \textit{et al.}}}(2003)}]{AHZ03}
\bibinfo{author}{\bibnamefont{{W. A. Atkinson \textit{et al.}}}},
  \bibinfo{journal}{Phys. Rev. B} \textbf{\bibinfo{volume}{68}},
  \bibinfo{pages}{054501} (\bibinfo{year}{2003}).

\bibitem[{\citenamefont{{G. Khaliullin \textit{et al.}}}(1997)}]{Ketal97}
\bibinfo{author}{\bibnamefont{{G. Khaliullin \textit{et al.}}}},
  \bibinfo{journal}{Phys. Rev. B} \textbf{\bibinfo{volume}{56}},
  \bibinfo{pages}{11882} (\bibinfo{year}{1997}).

\bibitem[{\citenamefont{{L. Zhu \textit{et al.}}}(2003)}]{ZAH03}
\bibinfo{author}{\bibnamefont{{L. Zhu \textit{et al.}}}},
  \bibinfo{journal}{Phys. Rev. B} \textbf{\bibinfo{volume}{67}},
  \bibinfo{pages}{094508} (\bibinfo{year}{2003}).

\end{thebibliography}
\end{document}